\documentclass[preprint,amsmath,amssymb,superscriptaddress,showpacs,aps12pt]{revtex4}

\usepackage[utf8]{inputenc}
\usepackage{graphicx}
\usepackage{epsfig}
\usepackage{amssymb}
\usepackage{dcolumn}
\usepackage{bm}

\newcommand{\ba}{\begin{array}}
\newcommand{\ea}{\end{array}}
\def\br{\begin{eqnarray}}
\def\er{\end{eqnarray}}
\def\be{\begin{equation}}
\def\ee{\end{equation}}
\usepackage{hyperref}

\def\({\left(}
\def\){\right)}

\def\S{\Sigma}
\def\<{\left\langle}
\def\>{\right\rangle}

\def\fc{\<{ \bar{\psi}}\psi\>}

\begin{document}

\title{Anomalous mass dimension of multi-flavor QCD}

\author{A. Doff}
\email{agomes@utfpr.edu.br}

\affiliation{Universidade Tecnol\'ogica Federal do Paran\'a - UTFPR - DAFIS
Av Monteiro Lobato Km 04, 84016-210, Ponta Grossa, PR, Brazil}

\affiliation{Instituto de F{\'i}sica Te\'orica, UNESP, Rua Dr. Bento T. Ferraz, 271, Bloco II, 01140-070, S\~ao Paulo, SP, Brazil}

\author{A. A. Natale} 
\email{natale@ift.unesp.br}

\affiliation{Instituto de F{\'i}sica Te\'orica, UNESP, Rua Dr. Bento T. Ferraz, 271, Bloco II, 01140-070, S\~ao Paulo, SP, Brazil}

\affiliation{Universidade Federal do ABC, Centro de Ci\^encias Naturais e Humanas,
Rua Santa Ad\'elia, 166, 09210-170, Santo Andr\'e, SP, Brazil}

\begin{abstract}
Models of strongly interacting theories with a large mass anomalous dimension ($\gamma_m$) provide an interesting possibility for the dynamical origin of the electroweak symmetry breaking. A laboratory for these models is QCD with many flavors, which may present a non-trivial fixed point associated to a conformal region.
Studies based on  conformal field theories and on Schwinger-Dyson equations have suggested the existence of bounds on the mass anomalous dimension at the fixed points of these models. In this note we discuss $\gamma_m$ values of multi-flavor QCD exhibiting a non-trivial fixed point and affected by relevant four-fermion interactions.  
\end{abstract}


%http://www.aip.org/publishing/pacs/pacs-reg10#11

\maketitle

%\section{Introduction}

Models of strongly interacting theories with a large mass anomalous dimension are important for
extensions of the standard model in the context of walking chiral symmetry breaking dynamics \cite{sannino,yamawaki}.  
These theories present a critical coupling constant for the onset of chiral symmetry breaking, related to a 
non-trivial fixed point associated to a conformal region.

Lattice QCD simulations provide a powerful tool to investigate the presence of a large anomalous dimension in the multi-flavor case. However, when the conformal behavior is approached as the number of flavor ($N_f$) is increased the simulations demand larger and larger lattice volumes. The enormous task to obtain limits on the anomalous dimension of multi-flavor QCD can be verified in Ref.\cite{aoki} and references therein.

Earlier determinations of these anomalous dimensions include a study based on Schwinger-Dyson equations (SDE) where it was verified that $\gamma_m \approx 1$ and is not strongly affected by high order corrections \cite{appelquist}. Another strong bound based on the unitarity of conformal field theories gives $\gamma_m \leq 2$\cite{mack}. Recent studies based on conformal bootstrap in $SU(N_f)_V$ symmetric conformal field theories suggest $\gamma_m < 1.31 $ for $N_f =8$ \cite{nakayama} and $\gamma_m \leq 1.29 $ for $N_f =12$ \cite{iha}. However the existence of a conformal fixed point in the $N_f=12$ model is still controversial \cite{fodor}.

It was observed for massless QED, when the coupling constant reaches a certain critical value, that the chiral symmetry is broken by the vacuum\cite{mira1}. Latter it was verified that at this critical coupling four-fermion interactions acquire an anomalous dimension such that the theory becomes renormalizable \cite{bard1,bard2}. The importance of a large anomalous dimension in what is now known as gauged Nambu-Jona-Lasinio  models became clear in the works of Refs.\cite{yama1, yama2, mira2, yama3, mira3, yama4}. In these models two coupling constants enter into action: the gauge coupling ($\lambda$) and the 4-fermion one ($g$), and there is a critical line described by a combination of these couplings where the chiral symmetry is broken. At this critical line the dynamical fermion mass behaves as a slowly decreasing function with the momentum
\cite{takeuchi,kondo}, and not much different from what is expected in a theory with bare masses.

The problem of mass generation with an enhanced fermionic condensate along the critical line associated to a slowly running
coupling, i.e. near a non-trivial fixed point, in non-Abelian gauge theories started to be studied in
Refs.\cite{apel, ap4,ap5}. In the Ref.\cite{apel} the fixed point was obtained from the two-loop $\beta$ function for a $SU(N)$ theory with fermions in the fundamental representation. One analysis of this problem in the case of other groups and fermionic representations can be seen in Ref.\cite{sn2}. With these studies begin the search for strongly interacting theories with a critical behavior, like the one of the
gauged Nambu-Jona-Lasinio model, associated to a non-trivial fixed point and consequently slowly running coupling constant,
possibly in the Banks and Zaks scenario \cite{BZ} of a perturbative value for the critical coupling. However in these cases the
critical line starts depending on the number of fermions and on the dimension of the fermionic representations. At this point we should mention that one of the contributions of this work, compared with the previous  studies, is that  we expand the resolution of the conformal window near the critical line  expected to exist in QCD, because we will be computing the anomalous mass dimension also in  the case of a non-perturbative fixed point.

In this note we will not provide a stronger bound on the anomalous dimension than the ones discussed in the previous paragraphs. We consider a different approach where a dynamical fermion mass has a hard asymptotic behavior associated to a non-trivial fixed point and relevant four-fermion interaction. This is also expected due to the coalescence of the SDE solutions \cite{georgi} . The theory will be fully described by the critical coupling $g_c$, where $\beta(g_c)\approx 0$, and the effect of the four-fermion interaction together with this critical gauge coupling lead us to be near the critical line, and this is translated in the fact  that we necessarily have a hard fermionic self-energy.

The anomalous dimension governs the renormalization effects on the fermion condensate at one large scale $\Lambda$ as
\be
\fc_\Lambda = exp \left[ \int_\mu^\Lambda \frac{d\kappa}{\kappa} \gamma_m \left( \alpha (\kappa)\right)\right] \fc_\mu \, .
\label{eq1}
\ee
Assuming that the theory has a slowly (or almost constant) running coupling constant in the interval $\mu$ to $\Lambda$ \cite{holdom} we have the following scaling relation
\be
\fc_\Lambda \approx \eta \left(\frac{\Lambda}{\mu}\right)^{\gamma_m} \mu^{3} \, ,
\label{eq2}
\ee
where $\eta $ is a constant and $\mu$ is the characteristic scale of the theory. A large $\gamma_m$ will modify the asymptotic dynamics and alleviate some of the problems suffered by composite models of dynamical electroweak symmetry breaking \cite{holdom}.

The fermion condensate as a function of the fermion self-energy ($\S(p^2)$) can be written as
\be
\fc_\Lambda = - \frac{N_R}{4\pi^2} \int_0^{\Lambda^2} dx \frac{x\S(x)}{x+\S^2(x)} \, ,
\label{eq3}
\ee
where $N_R$ is the dimension of the fermionic representation. We can compute $\fc_\Lambda $ with Eq.(\ref{eq3}) and one
specific expression for $\S(p^2)$ and compare it with the infrared condensate value in order to obtain a value for $\gamma_m$. 

It is known that the infrared behavior of the fermionic self-energy solution is a constant up to the scale $\mu$ that roughly defines the infrared dynamically generated mass. Therefore, apart from constants that may be assimilated in the infrared mass scale $\mu$ we have
\be
\fc_\mu \propto \mu^3 \, .
\label{eq3a}
\ee 
This is what is expected in standard QCD no matter the self-energy is a soft one like
\be
\Sigma^{(1)} (p^2 \rightarrow \infty) \sim \frac{ \mu^3 }{p^2} \, ,
\label{eq4}
\ee
which is the known behavior predicted by the operator product expansion \cite{politzer}, or has a harder behavior than
this one. Eq.(\ref{eq3a}) is just reflecting the infrared condensate behavior.

If the theory is affected by relevant four-fermion interactions the self-energies may vary up to the hardest behavior
with the momentum, as shown in Ref.\cite{takeuchi,kondo}, and is equal to 
\be
\Sigma^{(0)} (p^2) \sim \mu \left[1 + bg^2 (\mu^2) \ln\left(p^2/\mu^2 \right) \right]^{-\delta} \, 
\label{eq5}
\ee 
where    
\be
\delta= 3c/16\pi^2 b,
\label{eq6}
\ee
and $c$ is the quadratic Casimir operator given by  
\be
c = \frac{1}{2}\left[C_{2}(R_{1}) +  C_{2}(R_{2}) - C_{2}(R_{3})\right] \, ,
\label{eq7}
\ee
where $C_2 (R_i)$ are the Casimir operators for fermions in the representations  $R_{1}$ and  $R_{2}$ that form a composite boson in the representation $R_{3}$. In the QCD case, for quarks in the fundamental representation $R=F$,  $c= [(N^2-1)/2N]$ and $b$ is the coefficient of $g^3$ term in the renormalization group $\beta$ function and is given by $b=(33-2N_f)/48\pi^2$. 

Eq.(\ref{eq5}) may be a legitimate non-perturbative self-energy solution driven by confinement when
the gluons have a dynamically generated mass, as discussed in Ref.\cite{us}, or is the solution when the theory is dominated by the four-fermion interactions \cite{takeuchi,kondo}, and also  can result of coalescence of the SDE solutions \cite{georgi}. The critical coupling for this occur is the same one considered in  Ref.\cite{appelquist}, determined earlier by the authors of Ref.\cite{alfa1,alfa2,alfa3,alfa4,alfa5,alfa6,alfa7} and equal to
\be
\alpha_c = \frac{\pi}{3c} \, .
\label{eq10}
\ee 
Note that effects like dynamical gluon mass generation or confinement have not been taken into account in Ref.\cite{alfa1,alfa2,alfa3,alfa4,alfa5,alfa6,alfa7}, however even in these cases we should expect a similar coupling constant critical value for the onset of chiral symmetry breaking \cite{us}.

Notice that Eq.(\ref{eq5}) is solution of the renormalization group equation, in Landau gauge, given by
\be
\left( \frac{\partial}{\partial \mu} + \beta \frac{\partial}{\partial g} \right)\S = \frac{3c}{8\pi^2}g^2 \S \, ,
\label{eq8}
\ee
and normalized such that $\S(p^2=0)=\mu$. If we go to a higher order in the coupling constant we would expect that the coefficient 
$bg^2(\mu^2)$ appearing  in Eq.(\ref{eq5}) would be traded by higher order terms of the $\beta$ function. It is interesting to note that 
Eq.(\ref{eq8}) can be a generalization of the solution shown in Eq.(\ref{eq5}), because it could be solved considering higher order terms in the
$\beta$ function.

Let us now consider the fermion condensate given by Eq.(\ref{eq3}) calculated at the high energy scale $\Lambda$ and
 $\S(p^2)$ given by Eq.(\ref{eq5}) . It is straightforward to
show that the leading contribution to the QCD fermion condensate is
\be
\fc_\Lambda \propto \frac{3}{4\pi^2} \Lambda^{2} \mu \left[ 1+bg^2(\mu) \ln{\frac{\Lambda^2}{\mu^2}} \right]^{-\delta} \, .
\label{eq11}
\ee
This is the extreme behavior for the condensate. With this result and Eq.(\ref{eq3a}) we obtain
\be
\frac{\fc_\Lambda}{\fc_\mu}= \frac{\Lambda^2}{\mu^2}\left[1+bg^2(\mu) \ln{\frac{\Lambda^2}{\mu^2}} \right]^{-\delta} \, .
\label{eq11a}
\ee
However
\be
\fc_\Lambda = Z_m^{-1} \fc_\mu \, ,
\label{eq11b}
\ee
and the renormalization constant $Z_m$ is related to the anomalous dimension as
\be
\gamma_m \equiv \mu \frac{\partial}{\partial \mu} \ln Z_m \, ,
\label{eq11c}
\ee
leading, with the help of Eq.(\ref{eq11a}) to
\be
\gamma_m(\Lambda^2) = 2-2\delta b \frac{g^2(\mu^2)}{\left[1+bg^2(\mu^2) \ln{\frac{\Lambda^2}{\mu^2}} \right]} \, .
\label{eq11d}
\ee

The anomalous mass dimension of Eq.(\ref{eq11d})  was calculated using the fermionic condensate Eq.(\ref{eq3}) and not directly from the gap equation as usual. Furthermore, we use the hard self-energy (Eq.(\ref{eq5})) and not the soft one giving by Eq.(\ref{eq4}). 

If we were to deal with a conventional theory (or standard QCD), the asymptotic freedom condition ($\alpha (\Lambda^2) \rightarrow 0$) applied to Eq.(\ref{eq11d}) leads to $\gamma_m =2$, where we recover the known result of Ref.\cite{mack}. However this is not the case we are interested in, which is the one of a large number of flavors, where the theory is dominated by four-fermion interactions and we have an almost conformal theory, i.e. $\beta(g)\approx 0$ (or strictly zero), and we cannot use the leading order coupling constant as
considered in Eq.(\ref{eq11d}).

\par Before making a detailed calculation of Eq.(\ref{eq11d}) taking into account a possible conformal behavior of the coupling constant, we can recall that at the same time that the chiral symmetry is broken and non-trivial self-energies like the one of Eq.(\ref{eq5}) are generated,  composite pseudo-scalar Goldstone bosons (the pions) and scalar massive bosons (the $\sigma$) are formed, whose wave functions (respectively $\Phi_{BS}^P$ and $\Phi_{BS}^S$) are solutions of the Bethe-Salpeter (BS) equation are related to the fermionic self-energy as
\be
\Sigma (p^2) \approx  \Phi_{BS}^P (p,q)|_{q \rightarrow 0} \approx \Phi_{BS}^S (p,q)|_{q^2 = 4 m_{dyn}^2 }\, .
\label{eq15}
\ee
These BS wave functions are subjected to a normalization condition \cite{man,les}. One analysis of this condition applied to the
pseudo-scalar and scalar boson wave functions given by Eq.(\ref{eq5}) can be found respectively in the Refs.\cite{lane1} and \cite{us2}.
We will not enter into details of the calculation but just present the final constraint, which is
\be
\delta > \frac{1}{2}\,.
\label{eq16}
\ee
In the QCD case Eq.(\ref{eq16}) imply $N_f > 5$. Gauge theories, with a not too large number of fermions, can feature a non-trivial
fixed point in or near the so-called ``conformal window"\cite{cw}.  Recently, the authors in Refs.\cite{ap2,ap3}  demonstrated using lattice simulations, that the conformal window for the $SU(3)$ gauge theory lies in the range $8 < N_{fc} < 12$. In this region we indeed have a slowly running coupling ($\beta\approx 0$), where the limit of Eq.(\ref{eq16}) is valid. Therefore in the subsequent calculations the limit imposed by Eq.(\ref{eq16}) will always be taken into account leading to a large number of fermions.

We now return to Eq.(\ref{eq11d}). This equation is a consequence of Eq.(\ref{eq5}), involves a determination of the coupling constant 
$g^2(\mu^2)$ in the large $N_f$ limit, in one region where $\beta (g)\approx 0$, when the theory is dominated by a fixed point and four-fermion interactions are relevant. However even if Eq.(\ref{eq5}) is a consequence of four-fermion interactions it is only dependent on the
leading coefficient ($b$) of the $\beta$ function, and all information about the walking behavior should be present in the coupling
constant. This means that this coupling, in order to disclose such behavior, should be computed perturbatively beyond leading order or within some non-perturbative approach. Both possibilities are going to be discussed in the following. 

We first consider the perturbative approach and follow the same steps of Ref.\cite{apel}, with a walking two-loop $\beta$ function given by
\be
\beta (\alpha_\mu) = - \alpha_\mu (\beta_0 \alpha_\mu + \beta_1 \alpha^2_\mu) \, ,
\label{eq18}
\ee
where $(\beta_0 \alpha_\mu + \beta_1 \alpha^2_\mu) <<1$ in order to warrant a walking behavior, as well as this quantity is positive in
order to assure asymptotic freedom. We thus have the following approximation for the coupling constant
\be
\alpha (p) \approx \alpha_\mu \left( 1- \frac{\beta(\alpha_{\mu})}{\alpha_\mu} \ln \frac{p}{\mu}\right) \, ,
\label{eq19}
\ee

\noindent
which, if inserted into Eq.(\ref{eq11d}), entails
\be
\gamma_m (\Lambda^2) \approx 2 - \gamma_m  +  \frac{1}{2}\frac{\beta [\alpha(\mu)]}{\alpha_c} \ln \frac{\Lambda^2}{\mu^2} \, ,
\label{eq20}
\ee
where $\gamma_m  = \frac{1}{2}\frac{\alpha (\mu^2)}{\alpha_c }$. 

Eq.(\ref{eq20}) is justified when the coupling constant is given by Eq.(\ref{eq18}), but small differences will appear as we go
to more loops in the $\beta$ function. Within this approximation and assuming $\beta(\alpha_\mu = \alpha_c)\approx 0$ we
obtain $1\lesssim \gamma_m (\Lambda) < 2$ which is the known range for this quantity \cite{sannino,yamawaki}. Note that the
complete independence of $\gamma_m$ on the scale happens only in the presence of a fixed point, if this is not the case
its perturbative determination will be also dependent, above the two-loop order of the $\beta$ function, on the 
renormalization scheme.

It is interesting to comment on the behavior of Eq.(\ref{eq11d}) in the case where we indeed have a fixed point. We can indicate
the value of the coupling constant as $\alpha^*$ and consider $\beta (\alpha_{\mu} = \alpha^*)=0$, and now the anomalous dimension
will be described by
\be
\gamma_m \approx 2 - \frac{1}{2}\frac{\alpha^*}{\alpha_c}\approx 2 - \gamma^* \, ,
\label{eq21}
\ee
where $\gamma^*\equiv\frac{1}{2}(\alpha^*/\alpha_c)$. We first consider the $\alpha^*$ as the coupling constant determined
perturbatively in the Banks and Zaks scenario (BZ) \cite{cw,sn2,mira}, where 
\be
\alpha^*=-4\pi \frac{11N-4N_f d(R)}{34N^2-2N_fd(R)[10N+6C_2(R)]} \, ,
\label{eq22}
\ee 
and $d(R)$ is the dimension of the representation $R$ of the group $SU(N)$. Assuming the QCD conformal window in the range 
$8 \lesssim N_{f} \lesssim 12$, for $N_f \lesssim 12$ and $\alpha^*$ given by Eq.(\ref{eq22}) we obtain an upper limit of 
$\gamma_m \lesssim 1.5$.

In the table I we  show the results obtained for $(\gamma_m)$, $\alpha*$ and $\gamma^{*}$, for QCD assuming $\beta(\alpha(\mu))= 0$ at four loops, as a function of $N_f$ in the $\overline{MS}$ scheme \cite{4l} in the range $(8 \lesssim  N_{f} \lesssim 12)$
\begin{table}[t]
\centering
\caption{Anomalous dimensions evaluated at the fixed points obtained from $\beta(\alpha(\mu))= 0$ at four loops in the range $(8 \lesssim N_{f} \lesssim 12)$,  where $\alpha^{*}$ is the coupling constant value at the fixed point for each $N_{f}$.}
\label{parset}
\begin{tabular*}{\columnwidth}{@{\extracolsep{\fill}}llll@{}}
\hline
%\\
$N_f$ & $\alpha^{*}$ & $\gamma^{*}$  & $\gamma_m$\\
%      &              &               &        \\
\hline
$8$   &  $1.138$    & $0.724$   &  $1.275$ \\ 
$9$   &  $0.952$    & $0.608$   &  $1.394$ \\ 
$10$  &  $0.445$    & $0.283$   &  $1.717$ \\ 
$11$  &  $0.259$    & $0.165$   &  $1.835$\\ 
$12$  &  $0.176$    & $0.112$   &  $1.887$\\
\hline
\end{tabular*}
\end{table}
\par
%\vspace*{1cm} 

As a second determination of $\gamma_m$ we consider the case where the fixed point is a result of the dynamical generation
of gluon masses (DGM) \cite{cornwall, cpb, bp, abp}. As a consequence of dynamical gluon mass generation ($m_g(k^2)$) by the QCD vacuum the theory acquires a non-trivial infrared (IR) fixed point\cite{anp}. The running coupling constant becomes IR finite at moderately small values 
\cite{jn1} and  we have the following non-perturbative $\beta$ function\cite{cp, cornwall2, cornwall3}
\be
\beta_{DGM}=-bg^3\Big(1-\frac{4 m_g^2}{\Lambda^2}e^{-\frac{1}{bg^2}}\Big),
\label{eq:betacornwall}
\ee
where  $b$ is the coefficient of $g^3$ term in the renormalization group $\beta$ function and, for simplicity, the running behavior of the dynamical gluon mass was neglected. The variation of $m_g$ with $N_f$ was determined in lattice simulations\cite{abbcq} and with Schwinger-Dyson equations \cite{abp2} for a small number of flavors. These results were extrapolated for a large number of flavors as in Refs.\cite{brr, rdn, jn2}, and we use the following parameterization $m_g^{-1}=m_{g_0}^{-1}.e^{0.05942 N_f}$ with $m_{g_0}= 440$MeV \cite{jn2}. The values of $\alpha^*$ for $N_f= 8, 9$, $10$ and $12$ can be obtained from the Table II of Ref.\cite{jn2} and the anomalous dimensions that can be obtained in this approach are shown in Table II

In this note we discussed the anomalous mass dimension of multi-flavor QCD. The calculation is performed in the case that the theory has
a non-trivial fixed point or critical line and the theory is dominated by four-fermion
%\par 
\newpage
\begin{table}[t!]
\centering
\caption{Anomalous dimensions evaluated at the fixed points obtained from the non-perturbative (DGM) $\beta(g)$ function \cite{jn2},  
in the range $(8 \lesssim  N_{f} \lesssim 12)$,  where $\alpha^{*}$ is the coupling constant value at the fixed point for each $N_{f}$.}
\label{parset1}
\begin{tabular*}{\columnwidth}{@{\extracolsep{\fill}}llll@{}}
\hline
%\\
$N_f$ & $\alpha^{*}$ & $\gamma^{*}$  & $\gamma_m$\\
%      &              &               &           \\
\hline
$ 8$   &  $0.640$    & $0.407$ &  $1.593$ \\ 
$ 9$   &  $0.700$    & $0.446$ &  $1.554$ \\ 
$10$   &  $0.780$    & $0.496$ &  $1.504$ \\ 
$12$   &  $1.060$    & $0.676$ &  $1.324$\\ 
\hline
\end{tabular*}
\end{table}

\noindent interactions leading naturally to a hard asymptotic quark self-energy\cite{takeuchi,kondo}, what may also happens due to the coalescence of the SDE solutions \cite{georgi} or due to a confinement mechanism of chiral symmetry breaking \cite{us}.

In the context of the simple BZ scenario at the two-loop level and within the conformal window with $N_f \lesssim 12$ we obtain $\gamma_m \lesssim 1.5$. Similar results are obtained in the case of fixed points obtained from $\beta(\alpha^{*}) = 0 $ at four loops,  for $N_f = 8-10$ and in the DGM non-perturbative case. The data set presented in  tables I and II points to the existence of a limit on the  mass anomalous dimension compatible with $\gamma_m \approx (1.3 - 1.5)$. Therefore we obtained new estimates of the mass anomalous dimension of multi-flavor QCD in the presence of critical coupling constants associated to perturbative and non-perturbative fixed points. In order to compare with previous  studies we can say that in this work we expand the resolution of conformal window near the critical line  expected to exist in QCD, see Ref.\cite{4l}, once  we were computing the mass anomalous dimension also in  the case of a non-perturbative fixed point. We hope that these $\gamma_m$  values may be confronted with  the ones that can be obtained with QCD simulations and conformal bootstrap methods.
%\newpage 

\section*{Acknowledgments}
This research was  partially supported by the Conselho Nacional de Desenvolvimento Cient\'{\i}fico e Tecnol\'ogico (CNPq), by grant 
2013/22079-8 of Funda\c{c}\~{a}o de Amparo \`{a} Pesquisa do Estado de S\~ao Paulo (FAPESP) and by
Coordena\c c\~ao de Aperfei\c coamento de Pessoal de N\'{\i}vel Superior (CAPES).

\end{document}